\documentclass[iop]{emulateapj}
\usepackage{hyperref}

\newcommand{\dfracns}{$D_{\rm frac}^{\rm N_2H^+}$}

\newcommand{\kms}{$\rm km~s^{-1}~$}
\newcommand{\kmsns}{$\rm{km\:s}^{-1}$}

\shorttitle{An Ordered Bipolar Outflow from a Massive Early-Stage Core}
\shortauthors{Tan et al.}

\begin{document}

\title{An Ordered Bipolar Outflow from a Massive Early-Stage Core}

\author{Jonathan C. Tan}
\affil{Depts. of Astronomy \& Physics, University of Florida, Gainesville, Florida 32611, USA\\jctan.astro@gmail.com}

\author{Shuo Kong}
\affil{Dept. of Astronomy, University of Florida, Gainesville, Florida 32611, USA}

\author{Yichen Zhang}
\affil{Departamento de Astronom\'ia, Universidad de Chile, Casilla 36-D, Santiago, Chile}

\author{Francesco Fontani}
\affil{INAF - Osservatorio Astrofisico di Arcetri, I-50125, Florence, Italy}

\author{Paola Caselli}
\affil{Max Planck Institute for Extraterrestrial Physics (MPE), Giessenbachstr. 1, D-85748 Garching, Germany}

\author{Michael J. Butler}
\affil{Max Planck Institute for Astronomy, K\"onigstuhl 17, 69117 Heidelberg, Germany}

\begin{abstract}
We present {\it ALMA} follow-up observations of two massive,
early-stage core candidates, C1-N \& C1-S, in Infrared Dark Cloud
(IRDC) G028.37+00.07, which were previously identified by their $\rm
N_2D^+$(3-2) emission and show high levels of deuteration of this
species. The cores are also dark at far infrared wavelengths up to
$\sim100\:{\rm{\mu}m}$. We detect $^{12}$CO(2-1) from a narrow,
highly-collimated bipolar outflow that is being launched from near the
center of the C1-S core, which is also the location of the peak
$1.3\:$mm dust continuum emission. This protostar, C1-Sa, has
associated dense gas traced by $\rm{C^{18}O}$(2-1) and DCN(3-2), from
which we estimate it has a radial velocity that is near the center of
the range exhibited by the C1-S massive core. A second outflow-driving
source is also detected within the projected boundary of C1-S, but
appears to be at a different radial velocity.
After considering properties of the outflows, we conclude C1-Sa is a
promising candidate for an early-stage massive protostar and as such
it shows that these early phases of massive star formation can involve
highly ordered outflow, and thus accretion, processes, similar to
models developed to explain low-mass protostars.
\end{abstract}


\section{Introduction}

Understanding how massive stars form is an important goal, since their
radiative, mechanical and chemical feedback play leading roles in
regulating the interstellar medium, star formation activity and
overall evolution of galaxies. Core Accretion (e.g., McKee \& Tan
2003, hereafter MT03) is one class of models for massive star
formation, which involve initial conditions of a high-mass,
self-gravitating starless core, followed by relatively ordered
collapse to a central disk and protostar (see Tan et al. 2014
for a
review). These models are scaled-up from those developed for low-mass
star formation (e.g., Shu et al. 1987), but, in the case of the MT03
Turbulent Core Model, involve nonthermal forms of pressure support,
i.e., turbulence and magnetic fields, for the initial core to be in
approximate pressure and virial equilibrium.

Alternatively, Competitive Accretion models 
(e.g., Bonnell et al. 2001; Wang et al. 2010) 
involve a massive star gaining most of its mass by competitive,
chaotic Bondi-Hoyle accretion in the center of a crowded protocluster
of mostly low-mass stars. In these models, the initial conditions of
massive star formation, i.e., the gas immediately surrounding the
protostar that is destined to become a high-mass star, do not involve
massive starless, self-gravitating cores, but rather low-mass cores,
with most of the mass reservoir joining later from the protocluster
clump.

To try and distinguish between these theories we have developed a
method for searching for massive starless and early-stage core
candidates. We first identify target regions in IRDCs using
mid-infrared, i.e., $8\:\rm{\mu}\rm{m}$ {\it Spitzer-IRAC}, extinction
(MIREX) mapping (Butler \& Tan 2009; 2012). We select regions that are
peaks in the resulting mass surface density, $\Sigma$, map. We further
check that these regions are dark in $24\:\rm{\mu}\rm{m}$ ({\it
  Spitzer-MIPS}) and $70\:\rm{\mu}\rm{m}$ ({\it Herschel-PACS})
images. We then search for $\rm{N_2D^+}$(3-2) emission with {\it
  ALMA}, since the abundance of this species, i.e., the deuteration
fraction \dfracns$\equiv{\rm{[N_2D^+]/[N_2H^+]}}$ is known to rise in
cold ($T<20\:$K), dense ($n_{\rm{H}}\gtrsim10^5\:{\rm{cm}^{-3}}$)
conditions, especially when CO molecules are largely frozen-out onto
dust grain ice mantles and the ortho-to-para ratio of $\rm{H_2}$ drops
to low values (e.g., Kong et al. 2015a). $\rm{N_2D^+}$ is known to be
a good tracer of low-mass starless cores that are on the verge of
collapse, i.e., pre-stellar cores (Caselli \& Ceccarelli 2012), as
well as early stage low-mass Class 0 sources (Emprechtinger et
al. 2009).

We carried out a pilot search of 4 IRDC regions
($\rm{C1},\:F1,\:F2,\:G2$) with {\it ALMA} in Cycle 0 (compact
configuration, 2.3\arcsec\ resolution), identifying
$6\:\rm{N_2D^+}$(3-2) cores (Tan et al. 2013, hereafter T13) by
projection of their $3\sigma\:\:\:l-b-v$ space $\rm{N_2D^+}$(3-2)
contours. The two most massive cores were in IRDC G028.37+00.07
(hereafter Cloud C) C1 region: we refer to these as C1 North and South
(C1-N, C1-S). We estimated masses in two ways: (1) from the MIREX map,
finding C1-N has $61\pm30\:M_\odot$ and C1-S has $59\pm30\:M_\odot$
with $\sim$50\% systematic uncertainty due to distance ($5\pm1\:$kpc)
and dust opacity ($\sim30\%$) uncertainties; (2) from mm dust
continuum emission, finding C1-N has $16^{33}_7\:M_\odot$ and C1-S has
$63^{129}_{27}\:M_\odot$, with uncertainties mostly due to the adopted
dust temperature of $T=10\pm3\:$K, together with distance and dust
emissivity uncertainties.

These {\it ALMA} observations resolve the cores with about three beam
diameters. C1-S appears quite round, centrally-concentrated and
monolithic, while C1-N shows evidence of multiple fragments. Given
their high level of deuteration (Kong et al. 2016) and their dark
appearance in {\it Herschel-PACS} images, even at wavelengths as long
as $100\:\rm{\mu}\rm{m}$, C1-S, and perhaps also C1-N, are amongst the
best known candidates of massive starless or early-stage cores.

However, we note \citet{2006ApJ...651L.125W} reported a water maser
detection in this area (just outside C1-S's lowest $\rm{N_2D^+}$(3-2)
contour), though at a different velocity ($59.5\:$\kmsns) and in
single channel ($0.66\:$\kmsns wide). This water maser was not
detected in the more sensitive observations of
\citet{2009ApJS..181..360C} and Wang et al. (2012). We also note that
\citet{2015A&A...577A..75P} have detected CO(8-7) and (9-8) emission
towards C1-N \& S with {\it Herschel-HIFI} ($\sim20\arcsec$
resolution) and argue that this emission results from turbulence
dissipating in low velocity shocks,
which could be either driven by large-scale turbulent motions from the
surrounding cloud or from protostellar outflow activity.


Here we search for potential protostars and outflow activity via
$^{12}$CO(2-1) and other tracers using an {\it ALMA} Cycle 2
observation. Below we describe the observations (\S\ref{S:obs}),
present our results (\S\ref{S:results}) and discuss their implications
(\S\ref{S:disc}).

\section{Observations}\label{S:obs}

We use data from our {\it ALMA} Cycle 2 project (2013.1.00248.S,
PI:Tan), which observed the C1 region in a compact configuration on
05-Apr-2015, yielding sensitivity to scales from $\sim10\arcsec$ to
$\sim1\arcsec$). The position of the field center was
R.A.=18:42:46.5856, Dec.=-04:04:12.361 (FK5 J2000 system)
($l=28.3230,\:b=+0.06750$). It was chosen to be between C1-N and C1-S,
slightly closer to C1-S. Thus both cores are within the 27\arcsec\ field
of view.

The spectral set-up included a continuum band centered at $231\:$GHz
with width $1.875\:$GHz, i.e., from $230.0625\:$GHz to
$231.9375\:$GHz. The achieved sensitivity was $0.045\:$mJy per
$1.51\arcsec\times0.84\arcsec$ beam. In this continuum band, each
channel has width $1.129\:$MHz, i.e., velocity resolution
$1.465\:$\kmsns. The $^{12}$CO(2-1) line frequency is
$230.538\:$GHz. C1-S's radial velocity from its $\rm{N_2D}^+$(3-2)
emission is $+79.40\pm0.01\:$\kmsns with 1D dispersion of
$0.365\:$\kmsns (i.e., FWHM$=0.860\:$\kmsns), so the sky frequency
of $^{12}$CO(2-1) from this source is $230.477\:$GHz. Thus
we are sensitive to the presence of $^{12}$CO(2-1) emission with
moderate velocity resolution.
Note, ambient CO molecules in the core and even the wider scale IRDC
are expected to be largely frozen-out onto dust grain ice mantles
(Hernandez et al. 2011). However, $^{12}$CO(2-1) emission near ambient
velocities is still likely to be very optically thick from this
region. In our data the 9 channels closest to the expected ambient
velocity of C1-S have velocities centered at:
$74.1,\:75.4,\:76.7,\:78.0,\:79.3,\:80.6,\:81.9,\:83.2,\:84.5\:$\kmsns.

The other basebands were tuned to observe $\rm{N_2D}^+$(3-2),
$\rm{C^{18}O}$(2-1), DCN(3-2), DCO$^+$(3-2), SiO(v=1)(5-4) and
$\rm{CH_3OH}\:{\rm{v}\:{t}}=0\:5(1,4)-4(2,2)$. These data will be
presented and analyzed in full in a future paper, while in this {\it
  Letter} we focus mostly on the results of the broad continuum band
and its $^{12}$CO(2-1) line, along with some results from
$\rm{C^{18}O}$(2-1) and DCN(3-2) that help probe denser gas of
protostellar cores.

\section{Results}\label{S:results}

\begin{figure*}[htb!]
\epsscale{1.18}\plotone{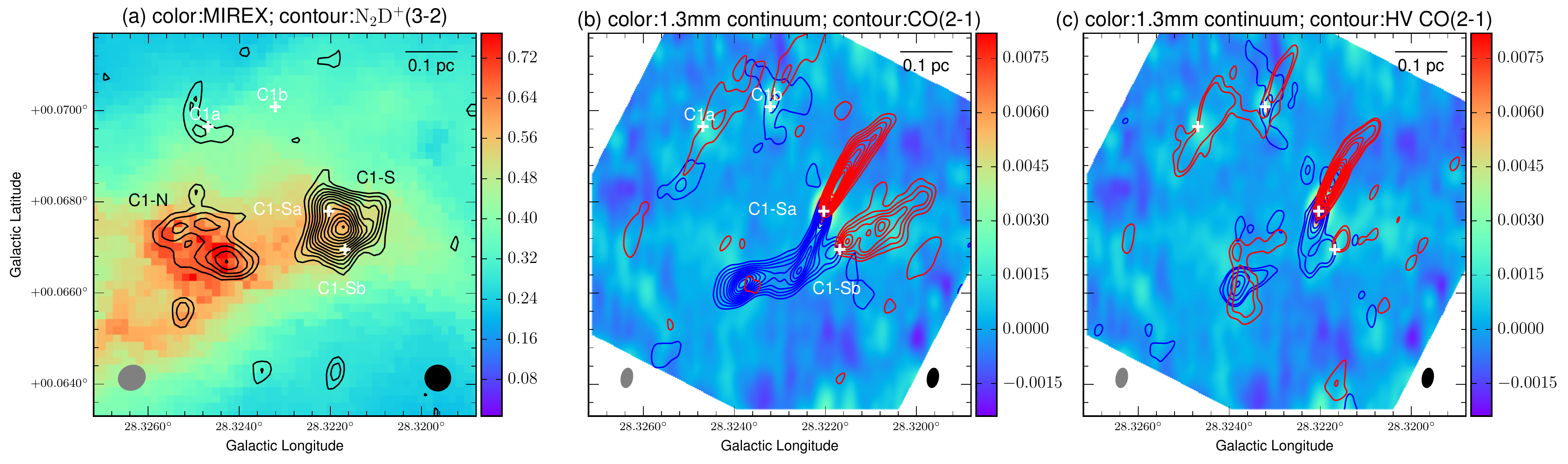}
\epsscale{1.18}\plotone{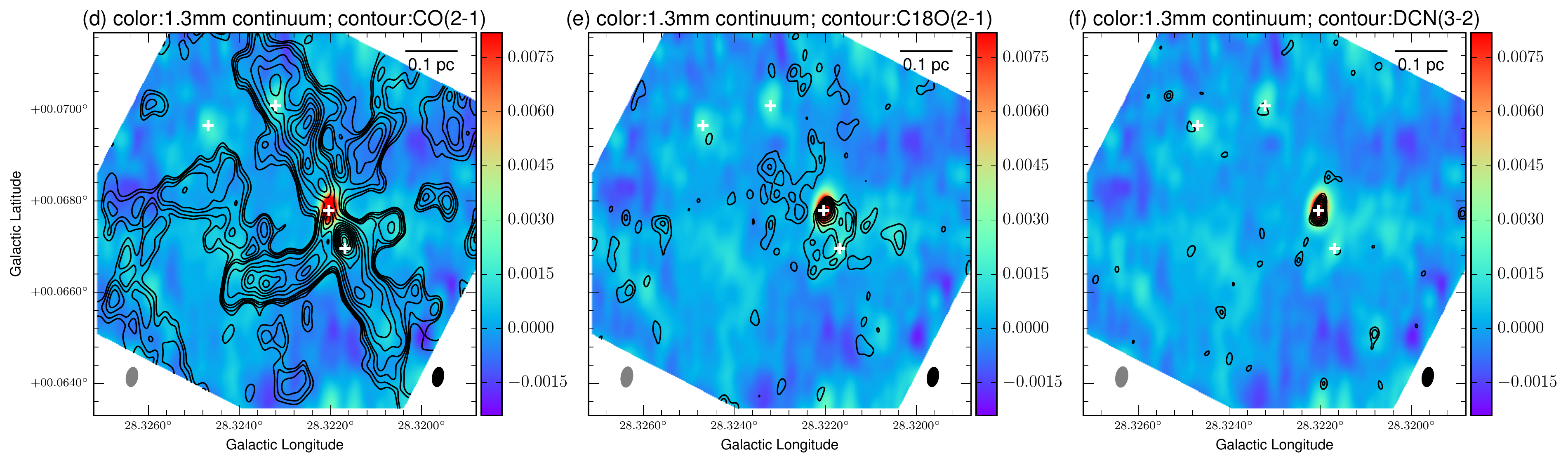}
\epsscale{1.18}\plotone{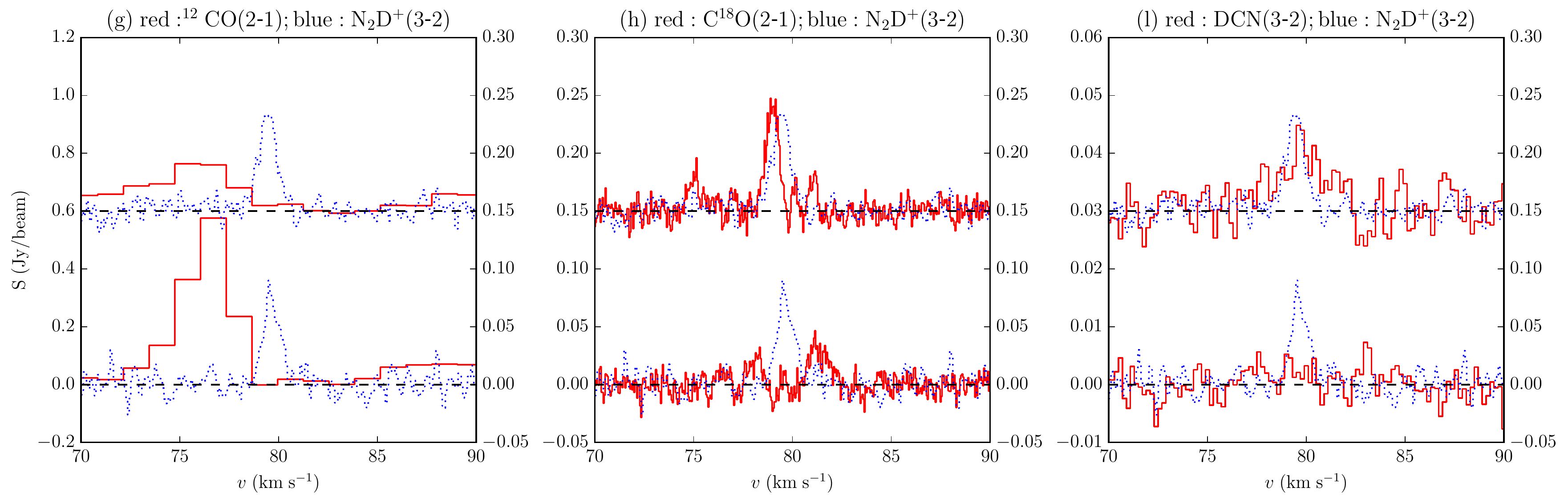}
\caption{
{\it (a) Top Left:} MIREX $\Sigma$ map (in $\rm{g\:cm}^{-2}$) of
C1 region, also showing $\rm{N_2D}^+$(3-2) emission (black contours)
observed with {\it ALMA} (Cycle 0) (T13) ({\it ALMA} beam is grey
ellipse in lower left; {\it Spitzer} beam that sets MIREX map
resolution is in lower right). $\rm{N_2D}^+$(3-2) contours are shown
from $2,\:3,\:4\sigma$..., with
$1\sigma\simeq10\:{\rm{mJy}\:{beam}^{-1}}\:$\kmsns. The C1-S core is
prominent at center-right, while the more fragmented C1-N core is
center-left.  Protostar candidates (``+'' symbols) are based on
mm-continuum peaks (panel b).
{\it (b) Top Middle:} $231\:$GHz continuum emission (color scale in
Jy/beam) from {\it ALMA} Cycle 2 observation with 1.2\arcsec\ beam
shown in lower left. Red contours show integrated intensity of
$^{12}$CO(2-1) from $v_{\rm{LSR}}=85.8\:{\rm{to}}\:124.8\:$\kmsns; blue
contours show emission from $33.8\:{\rm{to}}\:72.8\:$\kms. Contour levels start
from $30,\:60,\:90\sigma$..., where
$\sigma=11.6\:{\rm{mJy}\:{beam}}^{-1}\:$\kmsns. Beam at line frequency
is at lower right.
{\it (c) Top Right:} Same as (b), but now only showing high velocity
(HV) gas that is $+20\:$\kmsns or greater (red contours) and
$-20\:$\kmsns or less (blue contours) than $v_{\rm{LSR}}$ of C1-S. The
contour levels are shown from $20,\:30,\:60,\:90\sigma$..., where
$\sigma=9.3\:{\rm{mJy}\:{beam}}^{-1}\:$\kmsns.
{\it (d) Center Left:} As (b), but now contours show ``ambient''
$^{12}$CO(2-1) integrated intensity over velocities
$75.4\:{\rm{to}}\:83.2\:$\kmsns, ranging from
$3,\:10,\:20,\:30,\:60,\:90\sigma$..., with
$1\sigma=5.9\:{\rm{mJy}\:{beam}}^{-1}\:$\kmsns.
{\it (e) Center:} As (b), but now contours show
$\rm{C^{18}O}$(2-1) integrated intensity over velocities
$74\:{\rm{to}}\:84\:$\kmsns, ranging from
$3,\:5,\:7,\:9,\:11,\:13,\:15,\:18\sigma$, with
$1\sigma=8.5\:{\rm{mJy}\:{beam}}^{-1}\:$\kmsns.
{\it (f) Center Right:} As (b), but now contours show DCN(3-2)
integrated intensity over velocities $77\:{\rm{to}}\:81\:$\kmsns,
ranging from $3,\:4,\:5,\:6,\:7,\:8,\:9\sigma$, with
$1\sigma=3.8\:{\rm{mJy}\:{beam}}^{-1}\:$\kmsns.
{\it (g) Bottom Left:} Spectra of $^{12}$CO(2-1) (red solid lines) and
$\rm{N_2D}^+$(3-2) (dotted blue lines) extracted over 1 beam area
towards C1-Sa (top, offset up) and C1-Sb (bottom).
{\it (h) Bottom Middle:} As (g), but showing $\rm{C^{18}O}$(2-1) and $\rm{N_2D}^+$(3-2).
{\it (i) Bottom Right:} As (g), but showing DCN(3-2) and $\rm{N_2D}^+$(3-2).
\label{fig:outflow}}
\end{figure*}

\begin{deluxetable*}{cccccc}
\tabletypesize{\scriptsize}
\tablecolumns{6}
\tablewidth{0pc}
\tablecaption{Cores and Protostars in C1 Region\label{tab:coremodel}}
\tablehead{
\colhead{Name} & \colhead{$l\:$($^\circ$)} & \colhead{$b\:$($^\circ$)} & \colhead{$S_{\rm{1.3mm}}\:$(mJy)} & \colhead{$v_{\rm{LSR}}\:$(\kmsns)} & \colhead{P.A.$_{\rm{outflow}}\:$($^\circ$)}
}
\startdata
C1-N & 28.32503 & 0.06724 & $6.94\pm0.72$ & $81.18\pm0.03$\tablenotemark{a} & ...\\
C1-S & 28.32190 & 0.06745 & $26.7\pm0.77$ & $79.40\pm0.01$\tablenotemark{a} & ...\\
\hline
C1-Sa & 28.322093 & 0.067698664 & $19.5\pm0.1$ & $79.01\pm0.12$\tablenotemark{b} & 155\\
C1-Sb & 28.321752 & 0.066847223 & $2.7\pm0.1$ & $81.36\pm0.42$\tablenotemark{c} & 113\\
C1a & 28.324765 & 0.069543149 & $3.6\pm0.1$ & $80.2$\tablenotemark{d} & 150\\
C1b & 28.323272 & 0.069987301 & $3.5\pm0.1$ & ... & 150
\enddata
\tablenotetext{a}{
$v_{\rm{LSR}}$ of C1-N and C1-S estimated from $\rm{N_2D}^+$(3-2)
  (T13). $v_{\rm{LSR}}$ of protostars estimated from strongest
  $\rm{C^{18}O}$(2-1) peak, but see also notes below for individual
  sources.
}
\tablenotetext{b}{
C1-Sa: $v_{\rm{LSR}}$ estimated from strongest $\rm{C^{18}O}$(2-1)
peak. Secondary peaks at $75.08\pm0.05\:$\kmsns,
$81.08\pm0.03\:$\kmsns, while DCN(3-2) has single peak at
$79.8\pm0.2\:$\kmsns.}  
\tablenotetext{c}{
C1-Sb: $v_{\rm{LSR}}$ estimated from strongest $\rm{C^{18}O}$(2-1)
peak. Secondary peak at $78.16\pm0.05\:$\kmsns. DCN(3-2) is too weak
to measure $v_{\rm{LSR}}$.}
\tablenotetext{d}{C1a: $v_{\rm{LSR}}$ tentatively estimated from weak
  $\rm{N_2D}^+$(3-2) emission (T13).}
\end{deluxetable*}

Figure 1a presents the MIREX $\Sigma$ map (Butler et al. 2014) of
the C1 region, together with contours of 
$\rm{N_2D}^+$(3-2) integrated intensity (T13), which define C1-S and
C1-N. Also shown are locations of potential protostars defined by
$1.3\:$mm dust continuum peaks detected in the {\it ALMA} Cycle 2
image (Fig.$\:1$b), which has been cleaned with natural weighting and
had a primary beam correction applied. Figure~1b also shows integrated
intensities of (continuum subtracted) $^{12}$CO(2-1) with red contours
tracing $v_{\rm{LSR}}=85.8\:{\rm{to}}\:124.8\:$\kmsns, i.e.,
redshifted velocities up to $45.4\:$\kms with respect to C1-S's
central velocity, and blue contours tracing from
$33.8\:{\rm{to}}\:72.8\:$\kmsns, i.e., blueshifted velocities up to
$45.6\:$\kmsns from C1-S. Figure~1c shows just high velocity outflow
gas that is $>20\:$\kmsns away from C1-S's ambient velocity. Figure~1d
shows the integrated intensity of ``ambient'' $^{12}$CO(2-1), i.e.,
from $75.4\:{\rm{to}}\:83.2\:$\kmsns. Figures~1e and 1f show
integrated intensities of $\rm C^{18}O$(2-1) and
DCN(3-2), respectively, both potentially helpful to identify dense gas
associated with protostellar cores (e.g., Parise et al. 2009), and
thus their radial velocities.

From these images we clearly identify two protostellar sources that
are spatially overlapped with the C1-S core (Table 1). We refer to the
more central source as C1-Sa and define its spatial position as the
location of the $1.3\:$mm continuum peak. This peak, with flux density
$11\:$mJy/beam, is 1.31\arcsec\ from C1-S's center as defined by
$\rm{N_2D}^+$(3-2) (T13). Recall that for an estimated (kinematic)
source distance of $5\pm1\:$kpc, 1\arcsec\ corresponds to $5000\:$AU,
i.e., $0.024\:$pc, with $\sim$20\% uncertainties. So the spatial
location of C1-Sa is quite close to center of the C1-S core, which has
a radius of 3.61\arcsec\ (i.e., $0.0875\:$pc;$\:18,000\:$AU).

Figures 1g, 1h and 1i show the spectra of $^{12}$CO(2-1),
$\rm{C^{18}O}$(2-1) and DCN(3-2), also in comparison with the T13
observation of $\rm{N_2D}^+$(3-2), towards the protostars.
We estimate the radial velocity of the C1-Sa protostar from the
$\rm{C^{18}O}$(2-1) and DCN(3-2) spectra towards the continuum
peak. The $\rm{C^{18}O}$(2-1) spectrum shows a main peak at
$+79.01\pm0.12\:$\kmsns, while DCN(3-2) shows a single peak at
$+79.8\pm0.2\:$\kmsns. Thus it seems very likely that C1-Sa is forming
inside the C1-S $\rm{N_2D}^+$(3-2) core, which has mean velocity of
$+79.4\:$\kms and FWHM of $0.86\:$\kmsns.

The Galactic coordinate frame position angle of C1-Sa's $^{12}$CO(2-1)
outflow axis, which we define to the blueshifted axis, is
$\simeq155^\circ$. The outflow is highly-collimated and is seen to
extend $\sim12\arcsec$ ($60,000\:$AU,$\:0.3\:$pc), and is quite
symmetric, i.e., P.A. of the redshifted lobe is almost 180$^\circ$
greater than that of the blueshifted lobe. Also the observed extent
of the outflow is similar in each direction, although the highest
velocity flow is more extended on the redshifted side. The outflow can
be traced down to 3$\sigma$ above the noise level,
without bunching of the contours, so we expect the observed extent is
simply due to observational sensitivity and the actual extent could be
much larger.

We re-checked our {\it ALMA} Cycle 0 data, which included a requested
bandpass set to an intermediate frequency between DCN(3-2) and
SiO($v=$0)(5-4). However, this frequency was later mistakenly shifted to
be closer to DCN(3-2) causing the SiO line center to be unobserved:
only the potential blue wing up to $v_{\rm{LSR}}=+69\:$\kms was
observed. For this reason, T13 did not report detection of any SiO
emission towards C1-S. However, now we do see indications of the blue
wing of SiO($v=$0)(5-4) overlapping with the central part of the blue lobe of
the $^{12}$CO(2-1) outflow and extending to
$v_{\rm{LSR}}\simeq+50\:$\kmsns. We conclude it is likely that the
outflow from C1-Sa also emits strongly in SiO($v=$0)(5-4) across its
full velocity range.

The second source, C1-Sb, has a much weaker $1.3\:$mm continuum flux
of $1.3\:$mJy/beam and is located 2.0\arcsec\ from the center of the
C1-S $\rm{N_2D}^+$(3-2) core and 3.3\arcsec\ from C1-Sa. The
$\rm{C^{18}O}$(2-1) spectrum towards C1-Sb shows a main peak at
$+81.36\pm0.42\:$\kmsns and a secondary peak (with about half the
equivalent width) at $+78.16\pm0.45\:$\kmsns. The DCN(3-2) spectrum
shows no particularly strong features, although a 3$\sigma$ peak is
seen in the integrated intensity map (Fig. 1f). We tentatively
assign C1-Sb's radial velocity to be that of the main
$\rm{C^{18}O}$(2-1) spectral feature. We discuss below that this
assignment is potentially supported by examination of the channel maps
of the $^{12}$CO(2-1) ``ambient'' gas.
If this radial velocity is correct, then it would suggest that C1-Sb
is not physically associated with the C1-S $\rm{N_2D}^+$(3-2) core,
and in fact may be part of a gas structure that is linked to the C1-N
core. However, we cannot exclude the possibility that C1-Sb is also
forming from the C1-S core.


The outflow from C1-Sb has a similar extent as that from C1-Sa and
appears to have a wider opening angle. It has a P.A.=113$^\circ$, and
on its blueshifted side the outflow spatially overlaps with that from
C1-Sa. Analysis of lower intensity contours (down to $3\sigma$)
indicates a bunching close to the higher intensity contours. This may
indicate that, unlike for C1-Sa, we are seeing the full extent of
C1-Sb's outflow.

In Figure~1b we identify two more candidate protostars that are away
from the C1-S and C1-N cores. C1a is located in a region with faint
$\rm{N_2D}^+$(3-2) emission with radial velocity of
$80.6\:$\kmsns. There is relatively faint $^{12}$CO(2-1) emission,
which may be driven from this source with P.A.$\sim130^\circ$. There
is no significant $\rm{C^{18}O}$(2-1) emission associated with this
source, and only a weak $3\sigma$ feature in the DCN(3-2) integrated
intensity map. C1b is close to C1a and has a similar P.A. of its
$^{12}$CO(2-1) emission of about $130^\circ$. There is no significant
$\rm{C^{18}O}$(2-1) or $\rm{N_2D}^+$(3-2) emission at C1b, and only a
very weak feature in DCN(3-2). Note, the protostellar nature of C1a
and C1b is uncertain, especially since some of the observed
$^{12}$CO(2-1) features may be affected by side-lobe contamination
from C1-Sa and C1-Sb.

\begin{figure*}[htb!]
\epsscale{1.0}\plotone{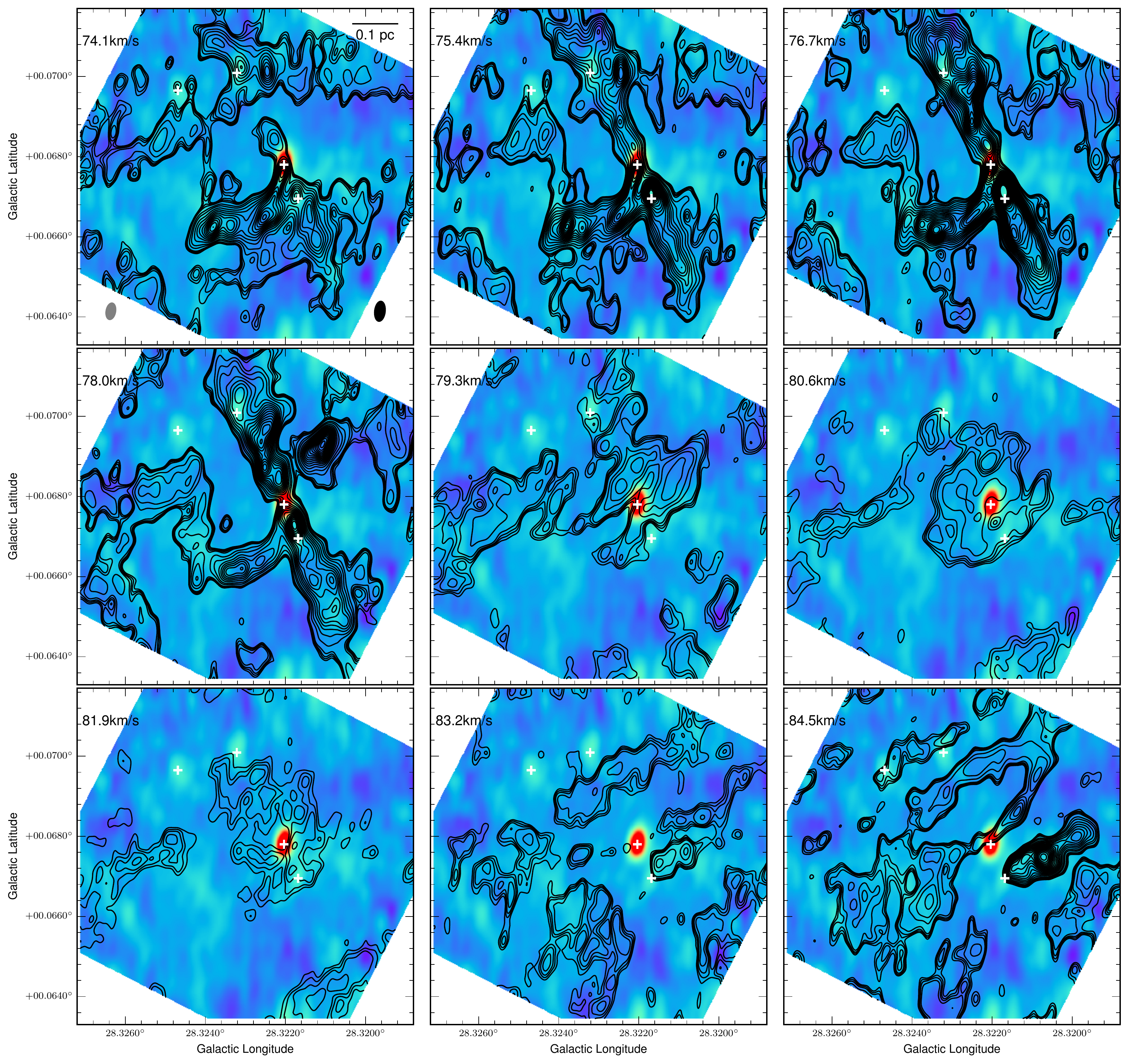}
\caption{
Channel maps of $^{12}$CO(2-1) emission, with contours starting from
$3,\:5,\:7,\:10,\:20,\:30,\:60,\:90\sigma$..., where
$\sigma=2.1\:{\rm{mJy}\:{beam}}^{-1}\:$\kmsns. Background image shows
$231\:$GHz continuum (Fig.~1b).
\label{fig:channel}}
\end{figure*}

Figure 2 shows nine channel maps of ``ambient'' $^{12}$CO(2-1)
emission from the C1 region, with high and low velocity ends
connecting with the ``outflow'' velocities plotted in Fig.~1b. The
C1-Sa outflow lobes are visible in the low and high velocity channels.
Fig.~2 also suggests the C1-Sb outflow has a driving source with a
$v_{\rm{LSR}}\simeq+81\:$\kmsns, since the blueshifted lobe is
already apparent in the $79.3\:$\kms channel, while the redshifted
lobe appears to vanish by the $81.9\:$\kms channel.

There are several other striking features seen in Fig.~2. First, there
is a very elongated ``filament'' that peaks in the 76.7 and
$78.0\:$\kmsns channels, but is visible from
$74.1\:\:\:{\rm{to}}\:\:\:79.3\:$\kmsns. The filament overlaps with the C1-Sa,
C1-Sb and C1b protostars and its orientation is almost perpendicular
to their outflows. The interpretation of this filament as an ambient
gas feature, rather than as a collimated bipolar outflow, is discussed
below considering its position-velocity diagram. Second, there is a
relatively weak, but still highly significant, quasi-spherical
``core'' of gas seen in the 80.6 and $81.9\:$\kmsns channels. Third,
there is $^{12}$CO(2-1) emission in the vicinity of the C1-N
$\rm{N_2D}^+$(3-2) core. Fourth, there are additional emission
features on the periphery of the image.

\begin{figure*}[htb!]
\epsscale{0.98} \plotone{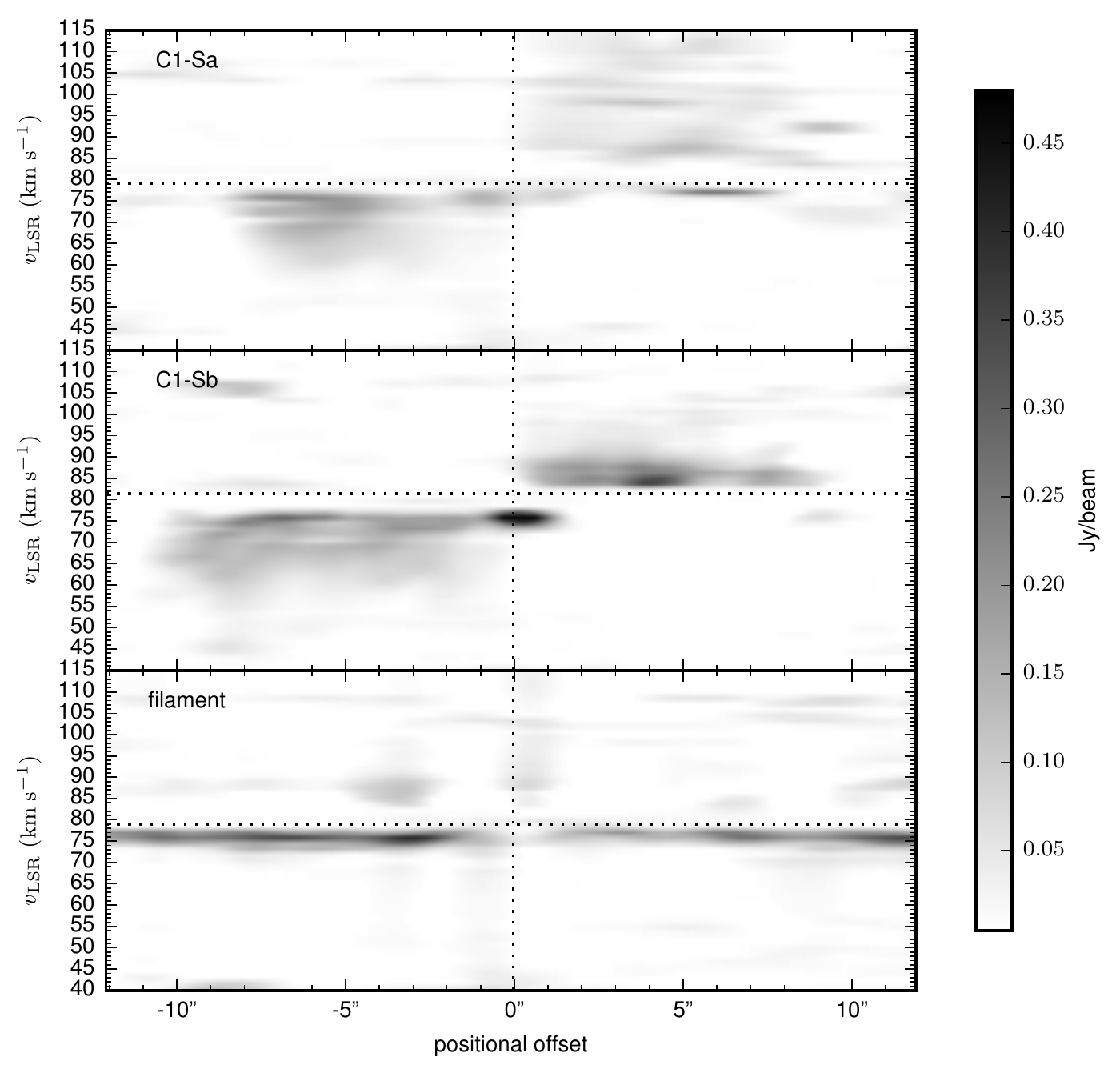}
\caption{
Position-velocity diagram of $^{12}$CO(2-1) emission along the axis of
the C1-Sa outflow (top), C1-Sb outflow (middle) and the ambient
``filament'' (bottom). Horizontal lines are included for reference at
$79.0\:$\kmsns for C1-Sa/filament and at $81.4\:$\kmsns for C1-Sb.
\label{fig:pv}}
\end{figure*}

Figure 3 shows position-velocity diagrams of $^{12}$CO(2-1) emission
along the outflow axes of C1-Sa and C1-Sb and along the axis of the
``ambient filament.'' These are defined by rectangular regions
3\arcsec\ wide running length-wise along the P.A. of each outflow or
the filament.

The ambient nature of the filament is readily apparent in these
figures. It has a very narrow velocity dispersion, which is comparable
to other ambient gas tracers, like $\rm{C^{18}O}$(2-1) and
$\rm{N_2D}^+$(3-2), and does not show a significant gradient in radial
velocity. The lack of such a gradient could be explained by an outflow
that was precisely aligned in the plane of the sky, but this would
still be expected to have a relatively broad velocity dispersion,
which the filament feature does not exhibit.

\begin{figure*}[htb!]
\epsscale{0.8}\plotone{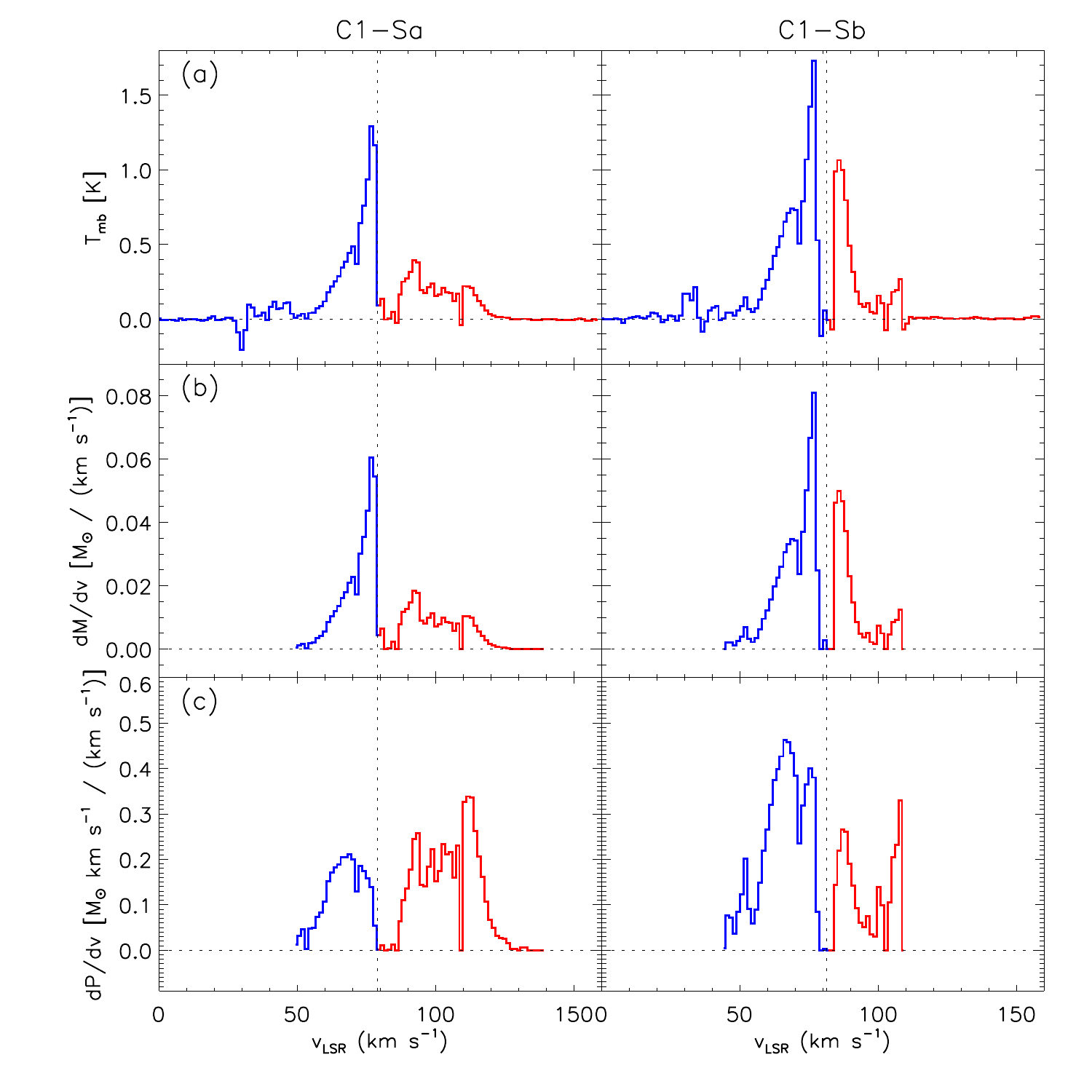}
\epsscale{0.8}\plotone{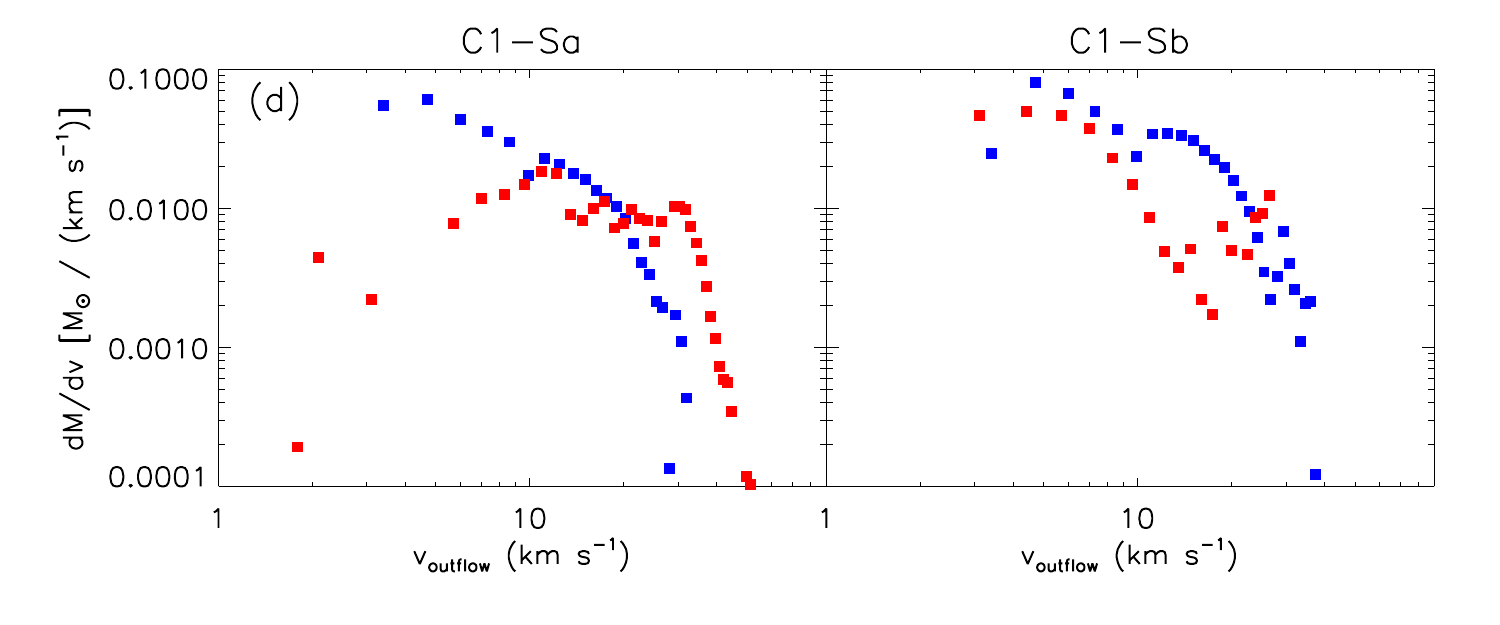}
\caption{
{\it (a) Top row}: $^{12}$CO(2-1) spectra of C1-Sa (left) and C1-Sb
(right), each extracted from a $24\arcsec\times3\arcsec$ aperture
centered on the protostars and aligned along the outflow axes. The
estimated ambient velocities of the protostars are shown with dotted
lines.
{\it (b) Second row:} Distribution of outflow mass (see text).
{\it (c) Third row:} Distribution of outflow momentum (see text).
{\it (d) Bottom row:} Comparison of mass distributions of blueshifted
and redshifted outflows versus outflow velocity, $v_{\rm outflow}$,
i.e., relative to the ambient velocity of each protostar.
\label{fig:outflow}}
\end{figure*}

Figure~4 shows $^{12}$CO(2-1) spectra and derived mass and momentum
distributions of the C1-Sa and C1-Sb outflows, extracted from the same
regions used for Fig.~3. To derive the mass, one must assume an
excitation temperature, with values of $T_{\rm{ex}}\simeq10$---$50\:$K
typically being used. $T_{\rm{ex}}=17\:$K minimizes the mass estimate,
while $50\:$K increases this by a factor of 1.5. 

For C1-Sa, integrating from $50<(v_{\rm{LSR}}/{\rm{km\:s}}^{-1})<140$
and assuming $T_{\rm{ex}}=17\:$K, the blue/red lobes have masses
$m_w=0.50/0.32\:M_\odot$
and momenta 
$p_w=3.6/6.6\:M_\odot\:$\kmsns,
respectively. Similarly, for C1-Sb, integrating from
$45<(v_{\rm{LSR}}/{\rm{km\:s}}^{-1})<110$ the blue/red lobes have
masses 
$0.73/0.38\:M_\odot$ 
and momenta 
$8.5/3.5\:M_\odot\:$\kmsns,
respectively. Note, the estimates for each blueshifted lobe are
affected by the overlap of the sources, leading to modest
overestimation of their properties, especially relative to the
redshifted lobes. However, in general the absolute values of the above
estimates should be viewed as lower limits, not only because of the
choice of $T_{\rm{ex}}=17\:$K, but also because of inclination effects
(which boost momentum estimates by $1/{\rm{cos}}(i)$, where $i$ is the
inclination of the outflow axis to the line of sight, with random
expectation value of $i=60^\circ$), and because of optical depth
effects (both within the outflowing gas, which may boost momentum by
factors of $\sim$6 (Zhang et al. 2016), and due to foreground
absorption).

For C1-Sa, the mass-weighted mean velocities ($v_w=p_w/m_w$) of the
blue/red lobes are 
$v_w=7.3/21\:$\kmsns, 
while for C1-Sb they are
$12/9.2\:$\kmsns.
Assuming the length of all the outflows lobes are
$\simeq12\arcsec\:(60,000\:$AU), the dynamical times for the blue/red
lobes of C1-Sa are 
$t_w=(3.9/1.3)\times10^4\:$yr 
and for C1-Sb are
$(2.4/3.1)\times10^4\:$yr. 
The correction factor for inclination is
cos($i$)/sin($i$), i.e., 0.58 for $i=60^\circ$. The mass outflow rates
for the C1-Sa blue/red lobes are
$(1.3/2.5)\times10^{-5}\:M_\odot\:{\rm{yr}^{-1}}$, 
and for C1-Sb are
$(3.0/1.2)\times10^{-5}\:M_\odot\:{\rm{yr}^{-1}}$. 
The correction factor for inclination is sin($i$)/cos($i$), i.e., 1.73
for $i=60^\circ$. Finally, the momentum injection rates for the C1-Sa
blue/red lobes are
$\dot{p}_w=(0.9/5.0)\times10^{-4}\:M_\odot\:$\kmsns$\:{\rm{yr}}^{-1}$
and for the C1-Sb blue/red lobes are
$(3.5/1.1)\times10^{-4}\:M_\odot\:$\kmsns$\:{\rm{yr}}^{-1}$. 
The inclination correction factor is sin($i$)/${\rm{cos}}^2i$, i.e.,
3.46 for $i=60^\circ$. Allowing for a mass underestimation factor of 3
and assuming $i=60^\circ$, the overall estimates of the total momentum
fluxes are boosted by a factor of 10, i.e., totals for C1-Sa of
$\dot{p}_w\sim5.9\times10^{-3}\:M_\odot\:$\kmsns$\:{\rm{yr}}^{-1}$ and
for C1-Sb of
$\sim4.6\times10^{-3}\:M_\odot\:$\kmsns$\:{\rm{yr}}^{-1}$.

\section{Discussion}\label{S:disc}

Outflow momentum flux is expected to be the most reliable direct
tracer of protostellar properties, since it should be independent of
the effects of ambient environment (unlike mass flux, which depends
mostly on the mass that has been swept-up by the primary
outflow). Models of massive protostar formation (Zhang et al. 2014)
based on the Turbulent Core Model (MT03) for cores of $60\:M_\odot$ in
a clump environment with
$\Sigma_{\rm{cl}}\simeq0.4\:{\rm{g\:cm}^{-2}}$ (relevant for C1-S)
predict
$\dot{p}_w\lesssim6\times10^{-3}\:M_\odot\:{\rm{km\:s^{-1}\:yr}^{-1}}$,
when the protostellar mass is $m_*\lesssim3\:M_\odot$, rising to
$\sim10^{-2}\:M_\odot\:{\rm{km\:s^{-1}\:yr}^{-1}}$ by the time
$m_*\simeq10\:M_\odot$. It is interesting that these estimates are
comparable with the observed values of $\dot{p}_w$ from C1-Sa and
C1-Sb. This suggests that, if C1-Sa is a massive protostar in the
process of formation, that it is currently at a very early stage,
i.e., has yet to accrete most of its mass. This would be broadly
consistent with the protostar having a relatively low luminosity such
that it does not appear yet as a MIR source. High angular resolution,
high sensitivity MIR to FIR observations, e.g., with {\it JWST}, are
needed to measure the SED of the protostar, which can then also
constrain protostellar models.
We conclude that we have detected protostars of relatively low current
masses. C1-Sa appears to be embedded within the C1-S massive, cold core,
as defined by $\rm{N_2D}^+$(3-2) emission. It thus has a large mass
reservoir from which to continue to grow: we speculate it is destined
to become a massive star. 

As traced by $^{12}$CO(2-1), C1-Sa's bipolar outflow is highly
collimated and has velocities extending to $\sim50\:$\kmsns. Similar
(blueshifted) velocities are seen in SiO($v$=0)(5-4) emission. Using
mass-weighted mean velocities, which are $\sim10\:$\kmsns, the
$i=60^\circ$ inclination-corrected outflow timescale is
$\sim2\times10^4\:$yr. However, since C1-Sa's outflow is likely to
extend to larger distances than we observe, this is probably a lower
limit on the duration of protostellar activity. Given the symmetric
and linear morphology of the outflow lobes, it appears that C1-Sa (and
C1-Sb) have not suffered significant dynamical disturbance from other
nearby (proto)stars during the period they have been driving these
outflows. This is consistent with assumptions of Core Accretion models
and is a constraint on Competitive Accretion models.

For constant instantaneous star formation efficiency from the core,
$\epsilon_{c}$, the fiducial Turbulent Core Model predicts
$m_*=\epsilon_{c}M_c(t/t_{*f})^2$, where the total star formation time
is
$t_{*f}\rightarrow1.29\times10^5(M_c/60\:M_\odot)^{1/4}(\Sigma_{\rm{cl}}/1\:{\rm{g\:cm}^{-2}})^{-3/4}\:$yr. So
for the $M_c=60\:M_\odot$ and
$\Sigma_{\rm{cl}}=0.4\:{\rm{g\:cm}^{-2}}$ case, then
$t>2\times10^4\:$yr implies $m_*>0.36\:M_\odot$ (assuming
$\epsilon_c\simeq1$, expected during early stages when outflow cavity
opening angles are small) and
$\dot{p}_w\gtrsim1\times10^{-4}\:M_\odot\:{\rm{km\:s^{-1}\:yr}^{-1}}$.



We conclude that C1-Sa is a good candidate for an early-stage massive
protostar and as such it shows that these early phases of massive star
formation can involve highly ordered outflow, and thus accretion,
processes (see also Zhang et al. 2015). The massive C1-S core is
potentially also forming a second protostar, C1-Sb: improved estimates
of its radial velocity is needed to clarify its association with this
core. The results presented complement work that has shown collimated
outflows can be launched from later-stage, more luminous massive
protostars (e.g., Beuther et al. 2002). They also illustrate that
there are similarities between high-mass protostars and their
lower-mass cousins forming from lower-mass cores.

\acknowledgments We thank Ke Wang and an anonymous referee for
comments and discussions that improved the paper. JCT/SK acknowledge
an NRAO/SOS grant and NSF grant AST1411527. PC acknowledges the
financial support of the European Research Council (ERC; project PALs
320620).  This paper uses {\it ALMA} data:
ADS/JAO.ALMA\#2013.1.00248.S; ADS/JAO.ALMA\#2011.0.00236.S.
{\it ALMA} is a partnership of ESO,
(representing its member states), 
NSF (USA) and NINS (Japan), together with NRC (Canada), NSC and ASIAA
(Taiwan), and KASI (Republic of Korea), in cooperation with the
Republic of Chile. The Joint {\it ALMA} Observatory is operated by
ESO, AUI/NRAO and NAOJ.


\end{document}